\documentclass[prd,onecolumn,nofootinbib,notitlepage]{revtex4-1}

\usepackage{amsfonts}
\usepackage{amsmath}
\usepackage{amssymb}
\usepackage{amsthm}
\usepackage{bm}
\usepackage{dcolumn}
\usepackage{epsfig}
\usepackage{graphicx}
\usepackage{graphics}
\usepackage[latin1]{inputenc}
\usepackage{latexsym}
\usepackage{rotating}
\usepackage{hyperref}

\usepackage[usenames]{color}

\begin{document}
\title{Einstein ?`sigue teniendo raz\'on?}

\author{Nicol\'as Yunes}
\affiliation{Department of Physics, Montana State University, Bozeman, MT 59717, USA}

\date{\today}


\maketitle



{\bf{Introducci\'on}}
\vspace{0.2cm}

El 25 de Noviembre de 1915, Albert Einstein le regala a la humanidad su \emph{Teor\'ia de la Relatividad General} (RG). Cien a\~nos despu\'es, la comunidad cient\'ifica continua regocij\'andose con las variadas consecuencias de esta teor\'ia, como la existencia de agujeros negros y estrellas de neutrones. Hoy por hoy, no existe prueba concreta que se\~nale que Einstein estaba equivocado, pero la b\'usqueda contin\'ua porque la RG es intr\'insicamente incompatible con la Teor\'ia de la Mec\'anica Cu\'antica. Las variadas hip\'otesis que intentan unificar estas dos teor\'ias  requieren una modificaci\'on fundamental de alguno de los pilares b\'asicos de una de las dos. La esperanza de muchos es que una futura observaci\'on encontrar\'a una desviaci\'on de la RG que se\~nalar\'a el camino correcto a seguir; tal vez as\'i se iniciar\'a una revoluci\'on cient\'ifica similar a la que ocurri\'o con la mec\'anica cl\'asica de Newton a ra\'iz de la observaci\'on de la precesi\'on del perihelio de Mercurio por Le Verrier en el a\~no 1859.

No es coincidencia, entonces, que haya habido un gran crecimiento en el \'area experimental dedicada a la RG desde su descubrimiento en 1915. Hist\'oricamente, los tests ``cl\'asicos'' de la teor\'ia, como la desviaci\'on de la luz estelar por el Sol, contribuyeron a  establecerla como la teor\'ia reinante. Desde entonces, se han realizado una inmensidad de experimentos terrestres, observaciones en el Sistema Solar, en otras galaxias y con p\'ulsares binarios, que contin\'uan confirmando la teor\'ia. Una parte de \'estos han refinado las comprobaciones cl\'asicas de la teor\'ia, encapsuladas hoy en el marco de la teor\'ia post-newtoniana parametrizada (ppN por sus siglas en ingl\'es)~\cite{Will:2014xja} de los a\~nos 70. Las pruebas modernas van mucho m\'as all\'a, y se concentran en la b\'usqueda experimental de desviaciones de los principios b\'asicos y fundamentales de la teor\'ia de Einstein. En este corto ensayo, intentaremos hacer un breve repaso de estas \'ultimas~\cite{Will:2014bqa,Will:2014xja,Psaltis:2008bb,Yunes:2013dva}. 


%

\vspace{0.5cm}

{\bf{Experimentos Terrestres y en el Sistema Solar}}. 
\vspace{0.2cm}

La mayor\'ia de los experimentos que se han realizado en nuestro planeta o cerca de \'el intentan verificar el \emph{Principio de Equivalencia}. En su versi\'on \emph{d\'ebil}, este Principio estipula que la aceleraci\'on de cualquier cuerpo libre producida por un campo gravitacional externo es independiente de su estructura y composici\'on, ignorando su auto-campo gravitacional. Actualmente, no existe indicio alguno que indique una violaci\'on de este principio. El l\'imite m\'as estricto ha sido encontrado por el grupo E\"ot-Wash con p\'endulos de torsi\'on~\cite{Schlamminger:2007ht}: la aceleraci\'on diferencial de dos cuerpos de distinta composici\'on debe ser menor que $2 \times 10^{-13}$ de la acelerac\'on promedio de ambos cuerpos. Otros experimentos, como MICROSCOPE, Galileo-Galilei, STEP y la medici\'on de la posici\'on de la Luna con l\'aseres, intentar\'an establecer mejores l\'imites~\cite{Will:2014bqa}. 

La versi\'on original de \emph{Einstein} de este Principio a\~nade la invariancia local de Lorentz y de posici\'on. Estos requerimientos exigen que los resultados de cualquier experimento local que no se desarrollen en un campo gravitatorio sean independientes de la velocidad y la posici\'on del laboratorio inercial donde sean realizados. El marco moderno en el cual se examina el principio de invariancia de Lorentz es el de la \emph{Extensi\'on del Modelo Est\'andar}~\cite{Mattingly:2005re}. En este marco, se agregan todo tipo de interacciones a la acci\'on del Modelo Est\'andar que violen la invariancia de Lorentz; luego, se intenta restringir la magnitud de los coeficientes de cada interacci\'on a trav\'es de experimentos con aceleradores de part\'iculas, o con otras observaciones, como la birrefringencia en la propagaci\'on de fotones, la radiaci\'on gravitacional de Cerenkov, y las oscilaciones de neutrinos.  

La invariancia local de posici\'on ha sido analizada a trav\'es de compariciones de relojes at\'omicos en diferentes campos gravitacionales. El l\'imite m\'as estricto impone desviaciones fraccionarias menores que $2 \times 10^{-4}$ y han sido obtenidos por Gravity Probe-A~\cite{1980PhRvL..45.2081V} con m\'aseres de hidr\'ogeno. El proyecto ACES planea mejorar este l\'imite con relojes de \'atomos atrapados en fr\'io, que ser\'an colocados en la Estaci\'on Espacial Internacional en 2016. Estos experimentos restringen la posibilidad de que las constantes fundamentales de la f\'isica no gravitacional cambien temporal y espacialmente.

La versi\'on \emph{fuerte} del Principio de Equivalencia tambi\'en se ha puesto a prueba con observaciones en nuestro Sistema Solar. Esta versi\'on le a\~nade a la de Einstein la inclusi\'on de experimentos que involucren a cuerpos muy densos que produzcan un campo gravitacional fuerte, como el de las estrellas. Una manifestaci\'on cl\'asica de la violaci\'on de este principio es el efecto de Nordtvedt~\cite{1968PhRv..169.1014N}, en el cual dos cuerpos estelares densos de distinta composici\'on se aceleran de manera diferente en presencia de un campo gravitacional externo. Este efecto es an\'alogo a una violaci\'on del principio de equivalencia d\'ebil, excepto que ahora los cuerpos en discusi\'on poseen su propia gravedad y la diferencia en aceleraci\'on es entonces proporcional a la energ\'ia gravitatoria de cada cuerpo. Este efecto no ha sido encontrado en ninguna observaci\'on, con el l\'imite m\'as severo impuesto a trav\'es de una combinaci\'on de mediciones de la posici\'on de la Luna con l\'aseres y de los experimentos del grupo de E\"ot-Wash~\cite{Will:2014xja}. 

Otra predicci\'on fundamental de la RG es que el movimiento o la rotaci\'on de los cuerpos con gravedad propia produce una correcci\'on en el campo gravitatorio, que es an\'aloga a un campo magn\'etico producido por una carga el\'ectrica en movimiento o por un dipolo magn\'etico. Por ejemplo, la rotaci\'on de la Tierra causa un arrastre del espacio-tiempo, que a su vez afecta la \'orbita de cualquier sat\'elite. En particular, este efecto causa que la l\'inea de nodos de la \'orbita de un gir\'oscopo y su momento angular de rotaci\'on sufran una precesi\'on, resultados conocidos como el \emph{efecto de Lense-Thirring} y la \emph{precesi\'on geod\'esica} respectivamente. Ambos efectos han sido verificados a trav\'es de observaciones de sat\'elites terrestres, como Gravity Probe B~\cite{Everitt:2011hp} y LAGEOS~\cite{Ciufolini:2004rq}. Estas observaciones han restringido desviaciones de las predicciones de la teor\'ia de Einstein con una precisi\'on del $0.3 \%$ y del $20\%$ respectivamente. 

\vspace{0.5cm}
{\bf{Tests con P\'ulsares Binarios}}. 
\vspace{0.2cm}

Todos los an\'alisis mencionados anteriormente tienen un factor en com\'un: utilizan campos gravitacionales d\'ebiles y pr\'acticamente est\'aticos, donde la curvatura del espacio-tiempo es muy peque\~na. La magnitud del campo gravitacional y de la curvatura del espacio-tiempo son proporcionales al potencial newtoniano (divido por el cuadrado de la velocidad de la luz $c$), $U = G M/(R c^{2})$ y al inverso del radio de curvatura ${\cal{R}}^{-1} = R^{-1} [G M/(R c^{2})]^{1/2}$, donde $G$ es la constante de Newton, mientras que $M$ y $R$ son la masa y el radio caracter\'isticos del sistema. Similarmente, la variabilidad temporal del campo gravitacional es proporcional a $v/c$, donde $v$ es la velocidad caracter\'istica del sistema. Por ejemplo, el sistema binario Sol-Tierra tiene $U \sim 10^{-8}$, ${\cal{R}}^{-1} \sim 10^{-12} {\rm{km}}^{-1}$ y  $v/c \sim 10^{-4}$. Es por ello que la astrof\'isica del sistema solar puede ser descripta por la mec\'anica cl\'asica de Newton con alta exactitud.   


Para verdaderamente poder someter a control la RG uno debe realizar experimentos donde la curvatura del espacio-tiempo y la magnitud y variabilidad temporal del campo gravitacional no sean extremadamente peque\~nos. Un ejemplo son los p\'ulsares binarios~\cite{Lorimer:2005bw}. Los p\'ulsares son estrellas de neutrones que rotan a altas velocidades, algunos a velocidades comparables a las cuchillas de una licuadora. Estas estrellas tienen grandes campos magn\'eticos ($\sim 10^{12} \; {\rm{G}}$) que aceleran fotones en un cono que rota r\'igidamente con la estrella, como un faro de luz. Si la luz del cono cruza el campo visual de la Tierra, los radiotelescopios terrestres registran un ``pulso de fotones,'' raz\'on  por la que son llamados p\'ulsares. Los p\'ulsares son extremadamente densos, con masas alrededor de $1.5 M_{\odot}$ y radios de s\'olo $\sim 12 \; {\rm{km}}$; sus campos gravitacionales y curvatura son enormes ($U \sim 0.2$ and ${\cal{R}}^{-1} \sim 0.04$/km) y la variabilidad temporal tambi\'en puede ser grande cuando se encuentran en sistemas binarios. El pulsar binario m\'as relativista que ha sido encontrado es el \emph{pulsar doble} J0737-3039A,B~\cite{Lyne:2004cj}, compuesto por dos estrellas de neutrones con per\'iodo orbital de $0.1$ d\'ias ($v/c \sim 10^{-3}$). Otros p\'ulsares binarios relativistas est\'an compuestos por una estrella de neutrones y una estrella enana blanca, por ejemplo J1738+0333, J1141-6545, y J0348+0432. La estrella de neutrones en este \'ultimo ejemplo tiene una masa de $\sim 2 M_{\odot}$ y es la m\'as masiva que ha sido encontrada. Recientemente, se ha hallado un sistema orbital triple, J0337+1715, que consiste de un p\'ulsar con dos estrellas enanas en \'orbita alrededor de \'el, y con per\'iodos orbitales de $1.6$ y $327$ d\'ias~\cite{Ransom:2014xla}.

A partir del descubrimiento de Hulse y Taylor~\cite{Hulse:1974eb} del sistema binario B1913+16 se han realizado muchos ex\'amenes de la RG con p\'ulsares binarios. De hecho, B1913+16 fue el p\'ulsar con el que se confirm\'o, por primera vez, que las \'orbitas de sistemas binarios compactos decaen a causa de la emisi\'on de ondas gravitacionales (ondulaciones en el espacio-tiempo producidas por la aceleraci\'on de cuerpos en movimiento) exactamente como lo predice la RG. Las verificaciones de RG con p\'ulsares binarios est\'an hoy encapsuladas en el marco del formalismo post-kepleriano parametrizado (ppK por sus siglas en ingl\'es)~\cite{Damour:1991rd}. En RG, los elementos orbitales en la parametrizaci\'on kepleriana (por ejemplo, el argumento del pericentro) adquieren correcciones relativistas que los fuerzan a evolucionar temporalmente. Esta evoluci\'on en RG est\'a determinada completamente por las masas del sistema binario. En los an\'alisis de ppK, uno construye curvas de la raz\'on de cambio de cada elemento orbital en el plano de las masas del sistema binario. Si las predicciones de RG son correctas, hay una \'unica intersecci\'on de todas las curvas. De esta manera, uno puede poner a prueba la hip\'otesis nula de que la RG es correcta, sin tener que especificar una teor\'ia alternativa determinada.     

Todos estas pruebas intentan confirmar, cada vez con mayor eficacia, el principio de equivalencia fuerte, restringiendo por ejemplo (i) la existencia del efecto de Nordtvedt, (ii) la existencia de sistemas de referencia especiales, (iii) la variabilidad de la constante de Newton, y (iv) la existencia de radiaci\'on gravitacional dipolar. Por ejemplo, la existencia de sistemas de referencia especiales ha sido restringida a trav\'es de la observaci\'on de las \'orbitas de p\'ulsares binarios y de su falta de torsi\'on en p\'ulsares aislados~\cite{Lorimer:2005bw}. Estas observaciones permiten imponer los l\'imites m\'as severos~\cite{Yagi:2013qpa} a la teor\'ia de Einstein-\AE{}ther~\cite{Jacobson:2000xp} (una teor\'ia que reintroduce un campo de \AE{}ther que selecciona un sistema de referencia especial) y a la gravedad ``khronom\'etrica''~\cite{Blas:2009qj} (el l\'imite de bajas energ\'ias de la gravedad de Ho\v{r}ava, una propuesta de gravedad cu\'antica que posiblemente es renormalizable). 

La existencia de la radiaci\'on dipolar merece un p\'arrafo aparte. En la RG, la radiaci\'on gravitacional dipolar no est\'a permitida debido a la conservaci\'on de momento lineal y al principio de equivalencia fuerte. Sin embargo, hay teor\'ias alternativas que no obedecen el principio de equivalencia fuerte, y por lo tanto, en donde la radiaci\'on dipolar esta permitida. Este tipo de radiaci\'on es mucho m\'as fuerte que la radiaci\'on gravitacional cuadrupolar de la RG, y entonces los sistemas binarios decaen mucho mas r\'apido en algunas teor\'ias alternativas. Las observaciones astrof\'isicas que han confirmado que los sistemas binarios decaen exactamente como lo predice la RG han permitido establecer l\'imites severos en ciertas teor\'ias escalar-tensoriales, como la de Brans-Dicke y las de $F(R)$~\cite{Will:2014bqa,Will:2014xja}. 

\vspace{0.5cm}
{\bf{Futuros Tests Astrof\'isicos}}. 
\vspace{0.2cm}

Hasta hoy, ning\'un test de RG con p\'ulsares binarios involucra una de las predicciones mas fant\'asticas de la teor\'ia de Einstein: los agujeros negros. Estos objetos son estrellas completamente colapsadas gravitacionalmente, con densidades tan altas que la fuerza gravitatoria no permite que la luz escape de sus \emph{horizontes de sucesos}. La existencia de agujeros negros en el universo ha sido comprobada indirectamente a trav\'es de la observaci\'on de las \'orbitas de estrellas tipo-S en el centro de la V\'ia L\'actea. Estas observaciones nos permiten deducir la existencia de un objeto compacto, Sag.~A*, en el centro de nuestra galaxia con una masa de $\sim 4 \times 10^{6} M_{\odot}$. Pero, a\'un hoy, no hemos podido determinar si este objeto compacto es realmente un agujero negro tal como los describen las soluciones de las ecuaciones de Einstein, o si es alg\'un otro tipo de objeto oscuro predicho por una teor\'ia alternativa. 

Los agujeros negros pueden ser fant\'asticos laboratorios para examinar la RG, ya que son objetos intr\'insicamente no lineales, cuyo campo gravitacional y curvatura del espacio-tiempo cerca del horizonte pueden llegar a la unidad. Ubicado a aproximadamente 8 kpc de nuestro Sistema Solar, Sag.~A* tiene un tama\~no angular de solo $\sim 10 \; \mu{\rm{s}}$, con lo cual su disco de acreci\'on a\'un no ha podido ser observado. La misi\'on principal del Event Horizon Telescope~\cite{Ricarte:2014nca} es precisamente resolver la sombra creada por el horizonte de Sag.~A* en su disco de acreci\'on. Para alcanzar esta meta, se utiliza interferometr\'ia de muy larga base, combinando datos de varios telescopios en el mundo. Estas observaciones nos permitir\'an medir el momento angular de rotaci\'on y el momento cuadrupolar de Sag.~A*, y as\'i verificar los teoremas de de agujeros negros ``sin pelo'' de RG. \'Estos predicen que los momentos multipolares del campo gravitacional exterior de los agujeros negros sin carga el\'ectrica dependen s\'olo de la masa y el momento angular de rotaci\'on. Otra manera de verificar estos teoremas es a trav\'es de observaciones de estrellas en \'orbitas bien cercanas a Sag.~A*. Estas \'orbitas dependen de todos los momentos multipolares de Sag.~A*, pero el efecto de estos decrece con el orden de los multipolos. La observaci\'on de una \'orbita con semi-eje mayor de menos de un miliparsec ser\'ia suficiente para medir el momento cuadrupolar de Sag.~A*~\cite{Psaltis:2014mca}.  

A\'un si los objetos compactos satisfacen estos teoremas, esto no implica necesariamente que la teor\'ia de Einstein haya sido verificada. Primero, las soluciones de Schwarzschild y de Kerr que describen agujeros negros en RG pueden tambi\'en ser soluciones en teor\'ias alternativas~\cite{Yunes:2013dva}. Esto depende de los detalles particulares de la teor\'ia en consideraci\'on; existen soluciones de agujeros negros en teor\'ias alternativas que no coinciden con las de Schwarzschild o Kerr~\cite{Yunes:2013dva}. Adem\'as, las verificaciones de RG con sombras de Sag.~A* son de hecho pruebas cuasi-estacionarias, ya que s\'olo dependen de las geod\'esicas que describen el movimiento de part\'iculas en el disco de acreci\'on. En particular, la din\'amica no lineal de las ecuaciones de Einstein no puede ser contrastada con estas observaciones. 

\vspace{0.5cm}
{\bf{Futuros Tests con Ondas Gravitacionales}}. 
\vspace{0.2cm}

Estos problemas pueden evitarse a trav\'es de la detecci\'on de ondas gravitacionales emitidas en los \'ultimos minutos de vida de sistemas binarios compactos (sean estrellas de neutrones o agujeros negros). Estos sistemas son ``limpios'' astrof\'isicamente hablando, ya que las incertidumbres astrof\'isicas tienen un impacto despreciable en la trayectoria de estos cuerpos y, por lo tanto, en las ondas gravitacionales emitidas~\cite{Yunes:2013dva}. Adem\'as, en los \'ultimos momentos de vida, estos sistemas binarios pueden adquirir velocidades muy altas ($v/c = 0.3$--$0.6$), de hasta dos \'ordenes de magnitud mayores que la velocidad orbital del p\'ulsar doble. Al poder acceder a campos gravitacionales enormes y altamente din\'amicos, con curvaturas gigantes del espacio-tiempo, uno podr\'a verificar la RG en un r\'egimen nunca antes explorado. Esto nos permitir\'a ir mucho mas all\'a del principio de equivalencia fuerte, y de los teoremas de agujeros negros sin pelo, explorando realmente la no linealidad de las ecuaciones de Einstein.    

?`Pero qu\'e aprenderemos realmente con la detecci\'on de ondas gravitacionales sobre la teor\'ia fundamental de la gravedad? Una de las predicciones b\'asicas de la RG es que las ondas gravitacionales tienen s\'olo dos polarizaciones, tal como las ondas electromagn\'eticas. En otras teor\'ias, las ondas pueden tener hasta seis polarizaciones diferentes (dos escalares, dos vectoriales y dos tensoriales) que afectan al espacio-tiempo de manera diferente~\cite{Will:2014bqa,Will:2014xja,Yunes:2013dva}. Por ejemplo, las ondas gravitacionales en teor\'ias escalares-tensoriales poseen tres polarizaciones. La detecci\'on de ondas gravitacionales nos permitir\'a evaluar esta predicci\'on y cualquier desviaci\'on ser\'ia catastr\'ofica para la RG. 

Otro principio b\'asico es que las ondas gravitacionales se propagan en el espacio-tiempo a la velocidad de la luz. Este hecho est\'a \'intimamente relacionado con la interpretaci\'on de que las ondas gravitacionales son \emph{gravitones}, part\'iculas elementales que no tienen masa, que se propagan en el espacio-tiempo. Algunas teor\'ias que intentan explicar la aceleraci\'on del universo reciente predicen que el gravit\'on tiene masa~\cite{deRham:2010kj} y, por lo tanto, que se propaga a una velocidad menor que la de la luz. En estas teor\'ias, la relaci\'on de dispersi\'on de gravitones masivos ser\'ia diferente a la de RG, lo cual afectar\'ia la evoluci\'on temporal de la fase de las ondas gravitacionales~\cite{Yunes:2013dva}.

Las ondas gravitacionales tambi\'en pueden limitar fuertemente violaciones de la simetr\'ia de Lorentz en el sector gravitacional, siempre y cuando las detectemos en coincidencia con ondas electromagn\'eticas que se originen de la misma fuente. Un ejemplo de esto ser\'ia la detecci\'on de ondas gravitacionales de la fusi\'on de estrellas de neutrones en un sistema binario y, coincidentemente, ondas electromagn\'eticas de la subsiguiente explosi\'on corta de rayos gamma~\cite{Nishizawa:2014zna}. Detecciones coincidentes de este estilo no son muy probables, pero una sola observaci\'on ser\'ia suficiente para imponer l\'imites mucho m\'as severos que los que existen actualmente~\cite{Hansen:2014ewa}.   

La detecci\'on de ondas gravitacionales nos permitir\'a explorar la estructura multipolar de la radiaci\'on en su totalidad, en vez de s\'olo limitar la existencia de la radiaci\'on dipolar. La evoluci\'on frecuencio-temporal de las ondas gravitacionales codifican no s\'olo la presencia de esta \'ultima, sino tambi\'en la existencia de modificaciones en la radiaci\'on cuadrupolar y de m\'as alto orden multipolar. Una verificaci\'on de las predicciones de la RG tendr\'ia implicaciones dr\'asticas en cuanto a la existencia de polarizaciones adicionales, ya que normalmente \'estas tambi\'en modifican el tipo de radiaci\'on gravitacional que es emitida por sistemas binarios.

Otro aspecto que podremos dilucidar con ondas gravitacionales es la estructura de la acci\'on de Einstein y Hilbert. Las teor\'ias de gravedad cu\'antica t\'ipicamente predicen modificaciones en la acci\'on, que luego introducen modificaciones en las ecuaciones de campo. \'Estas son generalmente correcciones cuadr\'aticas (o de m\'as alta potencia) en invariantes de la curvatura, que a su vez pueden modificar la estructura de los agujeros negros y de las estrellas de neutrones, como por ejemplo es el caso en la teor\'ia din\'amica de Chern-Simons~\cite{Alexander:2009tp} y la teor\'ia de Einstein-Dilaton-Gauss-Bonnet~\cite{Yunes:2011we}. Si uno requiere que estos modelos sean derivados en el l\'imite de bajas energ\'ias de la teor\'ia heter\'otica de las cuerdas, de teor\'ias efectivas de la inflaci\'on, o de la teor\'ia cu\'antica de bucles~\cite{Alexander:2009tp}, toda modificaci\'on ser\'ia suprimida por la distancia de Planck. Pero requerimientos de este tipo pueden ser peligrosos, ya que pueden existir mecanismos a trav\'es de los cuales predicciones derivadas de argumentos dimensionales terminan siendo incorrectas, como en el caso de la constante cosmol\'ogica. 

Las ondas gravitacionales tambi\'en podr\'an limitar fuertemente una clase espec\'ifica de teor\'ia escalar-tensorial que ha evadido controles  basados en observaciones en el Sistema Solar~\cite{damour_esposito_farese}. En estas teor\'ias, la gravedad se reduce efectivamente a la RG cuando la curvatura es d\'ebil. Pero en situaciones de gravedad extrema, como en las estrellas de neutrones, un campo escalar puede activarse y modificar la evoluci\'on de sistemas binarios a trav\'es de la emisi\'on gravitacional dipolar. Este \emph{mecanismo de ocultaci\'on} es opuesto a lo que a veces ocurre en cosmolog\'ia, donde las situaciones con gravedad fuerte se reducen a RG, pero las predicciones de la teor\'ia de Einstein son modificadas a grandes escalas~\cite{Khoury:2003rn}. La detecci\'on de ondas gravitacionales emitidas en la fusi\'on de estrellas de neutrones puede restringir la existencia de estos efectos de manera comparable a la observaci\'on de p\'ulsares binarios~\cite{Sampson:2014qqa}.    

Finalmente, otro principio b\'asico de la RG que podr\'a ser verificado con ondas gravitacionales es la existencia de horizontes de sucesos en agujeros negros. Cuando dos agujeros negros colisionan, el residuo es un nuevo agujero negro cuyo horizonte din\'amico~\cite{Ashtekar:2004cn} vibra como una membrana~\cite{membrane}, produciendo ondas gravitacionales con modos cuasi-normales. Estos modos pueden ser utilizados para confirmar los teoremas de agujeros negros sin pelo de RG, y espec\'ificamente la estructura del espacio-tiempo de los agujeros negros din\'amicos. A diferencia de las observaciones electromagn\'eticas con discos de acreci\'on, estas pruebas no se ven afectadas por incertidumbres astrof\'isicas.

Todos estas comprobaciones de RG con ondas gravitacionales han sido encapsuladas en el marco del formalismo parametrizado post-einsteiniano~\cite{Yunes:2009ke} (ppE por sus siglas en ingl\'es). En este formalismo, uno agrega ciertos t\'erminos a los filtros que ser\'an utilizados para detectar ondas gravitacionales de sistemas binarios con la t\'ecnica de \emph{filtros emparejados}~\cite{Jaranowski:2007pe}. Estos t\'erminos representan modificaciones en la evoluci\'on de la amplitud y la fase de las ondas gravitacionales de una meta-teor\'ia que es capaz de reproducir todas las predicciones de teor\'ias alternativas, incluyendo la RG. Usando m\'etodos estad\'isticos de \emph{Bayes}, se puede identificar la meta-teor\'ia que mejor se adapta a los datos experimentales u observacionales.  

Otro marco en el cual se pueden realizar tests de la RG con ondas gravitacionales es mediante el uso de las relaciones de \emph{I-Love-Q}~\cite{Yagi:2013awa}. Estas relaciones predicen que el momento de inercia ($I$), el n\'umero de Love (que determina la deformabilidad de las estrellas), y el momento cuadrupolar ($Q$) de las estrellas de neutrones est\'an inter-relacionados de una manera que es \emph{independiente} de la estructura interna de las estrellas, pero depende de la teor\'ia de la gravedad. Es posible que observaciones futuras de p\'ulsares binarios permitan medir el momento de inercia de las estrellas de neutrones. Es tambi\'en posible que la detecci\'on de ondas gravitacionales producidas en la colisi\'on de estrellas de neutrones permitan medir el n\'umero de Love. Por lo tanto, a trav\'es de las relaciones de I-Love uno puede verificar si estas observaciones son consistentes con las predicciones de RG, sin que importe la estructura interna de las estrellas. Este tipo de pruebas permitir\'an limitar severamente la existencia de ciertas teor\'ias efectivas de la gravedad cu\'antica~\cite{Yagi:2013awa}. 

\vspace{0.5cm}
{\bf{Caminante no hay camino}}. 
\vspace{0.2cm}

Una de las \'areas fundamentales de la ciencia es la falsificaci\'on de hip\'otesis y el derrumbamiento de estructuras te\'oricas para dar lugar a rascacielos cada vez m\'as inclusivos. Preguntarse por qu\'e seguir poniendo a prueba la teor\'ia de Einstein es como preguntarse por qu\'e intentar entender el funcionamiento del universo. No se puede avanzar en la ciencia, y en particular en la astrof\'isica, sin un continuo cuestionamiento de nuestros m\'etodos y nuestros resultados. No hacerlo ser\'ia establecer un dogma que estancar\'ia el progreso de la ciencia indefinidamente. En los \'ultimos cien a\~nos, la comunidad cient\'ifica ha realizado estos cuestionamientos de manera exhaustiva, y sin embargo, tal vez insospechadamente, la RG ha aguantado sistem\'aticamente todos estos ex\'amenes y todos los intentos de modificaci\'on. Pero a\'un as\'i queda mucho m\'as por hacer. ?`Son los objetos oscuros y compactos del universo realmente los agujeros negros de Einstein? ?`Los gravitones son masivos o no? ?`Es el universo sim\'etrico, como lo predijeron Lorentz y Einstein con su principio de equivalencia? Tal vez las estructuras fundamentales del universo son m\'as que un promedio de las fluctuaciones que s\'olo efectivamente pueden ser descriptas por un tensor m\'etrico. Estas preguntas y muchas m\'as podr\'an responderse en los pr\'oximos a\~nos a trav\'es de los incre\'ibles esfuerzos te\'oricos y experimentales que est\'an hoy reci\'en empezando. Es imposible adivinar el camino que la ciencia seguir\'a en el futuro cercano; s\'olo podemos esperar y seguir haci\'endonos camino al andar.  

\vspace{0.5cm}
{\bf{Agradecimientos}}. Me gustar\'ia agradecerle a Jos\'e Senovilla, a Irene Grimberg y a Mariana Goldman por leer este art\'iculo minuciosamente y sugerir correcciones. Tambi\'en quisiera agradecerle a Diego Blas por varias discusiones cient\'ificas al respecto de este art\'iculo. Finalmente, quisiera agradecerle a Clifford Will, quien me ha otorgado permiso para utilizar material de sus art\'iculos~\cite{Will:2014bqa,Will:2014xja} en este ensayo.

\vspace{0.2cm}

\bibliographystyle{apsrev}
\bibliography{review}

\end{document}